\documentclass[11pt,a4paper]{article}
\usepackage{jheppub}
\usepackage[T1]{fontenc}
\usepackage{amsmath}
\usepackage{amssymb}
\usepackage{esint}
\usepackage{amsfonts}
\usepackage{graphicx,color}
\usepackage{slashed}
\usepackage{young}
\usepackage[vcentermath]{youngtab}
\usepackage[utf8]{inputenc}
\usepackage{tensor}
\usepackage{etoolbox}
\usepackage{bbm}
\usepackage{soul}
\usepackage{amsthm,mathabx}
\usepackage{braket}

\makeatletter

\newcommand{\be}{\begin{equation}}
\newcommand{\ee}{\end{equation}}
\newcommand{\bea}{\begin{eqnarray}}
\newcommand{\eea}{\end{eqnarray}}

\renewcommand{\tilde}{\widetilde}

\def\vp{\varphi}
\def\pa{\partial}

\newcommand*\xbar[1]{%
  \hbox{%
    \vbox{%
      \hrule height 0.5pt 
      \kern0.3ex
      \hbox{%
        \kern-0.0em
        \ensuremath{#1}%
        \kern-0.0em
      }%
    }%
  }%
} 

\pdfpageheight\paperheight
\pdfpagewidth\paperwidth

\pdfoutput=1 

    \patchcmd{\maketitle}{\@fpheader}{}{}{}

\title{BV-BRST Noether theorem} \author[a]{Glenn Barnich,} \author[b]{Laurent
  Baulieu,} \author[a,c]{Marc Henneaux} \author[d]{and Tom Wetzstein}
\affiliation[a]{Universit\'e Libre de Bruxelles and International Solvay
  Institutes, ULB-Campus Plaine CP231, B-1050 Brussels, Belgium}
\affiliation[b]{LPTHE, Sorbonne Universit\'e, CNRS, 4 Place Jussieu, 75005
  Paris, France} \affiliation[c]{Coll\`ege de France, Universit\'e PSL, 11 place
  Marcelin Berthelot, 75005 Paris, France} \affiliation[d]{Okinawa Institute of
  Science and Technology, 1919-1 Tancha, Onna-son, Okinawa 904-0495, Japan}

\abstract {The BRST Noether theorem, or ``Noether's $1.5$ theorem'', asserts the
  triviality of the BRST Noether current. We provide two proofs of this theorem
  that are both valid without restriction on the structure of the gauge theory,
  extending thereby previous proofs holding in the case of gauge theories for
  which the solution of the master equation is linear in the antifields. We also
  relate explicitly the BRST Noether current to the BRST master current
  appearing in the master equation.}

\makeatother

\begin{document}
\maketitle \flushbottom

\section{Introduction}
\setcounter{equation}{0}

In recent papers \cite{Baulieu:2024oql,Baulieu:2024rfp}, a version of Noether's theorem lying somehow ``half-way''
between Noether's first theorem valid for global (rigid) symmetries and
Noether's second theorem valid for local (gauge) symmetries \cite{Noether:1918zz} was found to
play an essential role in the description of asymptotic symmetries. This
theorem, coined Noether's $1.5$ theorem in \cite{Baulieu:2024oql}, asserts that the conserved
current of the BRST symmetry of the gauge-fixed action is BRST-trivial, just as
the Noether currents of the gauge symmetries of the gauge-invariant action are
trivial in the ``characteristic cohomology'' (defined as the quotient of the space
of ``on-shell'' closed $p$-forms by the subspace of on-shell exact ones \cite{Bryant:1993,Anderson:1989}).

The theorem was proved for gauge theories of ``rank-1'', for which the solution of
the master equation \cite{Batalin:1981jr,Batalin:1983ggl,Batalin:1985qj} is linear in the antifields. It is the purpose of this
note to provide proofs of the theorem valid without restriction on the structure
of the gauge theory. In the course of the analysis, we introduce the ``BRST
master current'', which is a gauge independent current depending on the fields
and the antifields. It is defined through the local version of the master
equation. It coincides with the BRST Noether conserved current upon gauge
fixing.

Our paper is organized as follows. In the next section (Section {\bf \ref{Sec:NOETHER12}}), we
review Noether's first and second theorems. Section {\bf \ref{Sec:Noether15T}} provides a first
proof of Noether's $1.5$ theorem, based on Noether's first theorem applied to
the gauge fixed action, as well as on the BRST-ghost number algebra. In Section
{\bf \ref{Sec:MasterC}} we introduce the ``BRST master current'' through the local version of
the master equation and prove its triviality in cohomology. This result is used
in Section {\bf \ref{Sec:2nd15}} to derive another proof of Noether's $1.5$ theorem.
Finally, Section {\bf \ref{Sec:Conclusions}} is devoted to a few comments. Three appendices,
reviewing useful properties of Euler-Lagrange derivatives as well as some
aspects of the antifield formalism, complete our paper.

\section{Noether's first and second theorems}
\label{Sec:NOETHER12}
As a way of fixing our notations and conventions, we briefly review in this
section the two Noether theorems.

\subsection{Noether's first theorem}
\label{sec:noeth-first-theor}

We start with the classical action
\begin{equation}
  \label{eq:17}
  S_0[\phi]=\int\, L_0\ d^n x \, , \qquad d^n x = dx^0 dx^1 \cdots dx^{n-1}
\end{equation}
which depends on the classical fields $\phi^i$ and a finite number of their
derivatives (``local functional''). The fields can be bosonic or fermionic. The
Euler Lagrange derivatives are defined as
\begin{equation}
  \label{eq:3}
  \frac{\delta^L L_0}{\delta\phi^i}=(-\partial)_{(\mu)}\frac{\partial^L L_0}{\partial\phi^i_{(\mu)}}
  =\frac{\partial^L L_0}{\partial \phi^i}-\partial_\mu\big(\frac{\partial^L L_0}{\partial \phi^i_\mu}\big)+\dots. 
\end{equation}
where a summation is implied over the multi-index $(\mu)$ with a sign that depends
on the parity of the number of indices contained in $(\mu)$.

We consider for simplicity variations of the fields $\delta_Q\phi^i=Q^i[\phi]$ with
$\delta x^\mu = 0$ (as can always be assumed by redefinitions), which are trivially
extended to derivatives by using the property $[\delta_Q,\partial_\mu]=0$,
\begin{equation}
  \label{eq:19}
  \delta_Q =\partial_{(\mu)}Q^i\frac{\partial^L}{\partial \phi^i_{(\mu)}}=Q^i\frac{\partial^L}{\partial \phi^i}+\partial_\mu Q^i\frac{\partial^L}{\partial \phi^i_\mu}+\dots . 
\end{equation}
Under such variations, the Lagrangian transforms as
\begin{equation}
  \label{eq:21}
  \delta_Q L_0=Q^i\frac{\delta^L L_0}{\delta\phi^i}+\partial_\mu \theta^\mu_Q,\quad \theta^\mu_Q= Q^i\frac{\partial^L L_0}{\partial\phi^i_\mu}+\dots. 
\end{equation}
The explicit form of the additional terms in $\theta^\mu_Q$ in higher derivatives
theories are obtained through repeated integrations by parts and may be found
e.g.~in \cite{Anderson:1989}.

By definition, the variation of the Lagrangian under a global symmetry is a
total derivative, which yields Noether's first theorem that any global symmetry
implies the existence of a conserved current (and conversely),
\begin{equation}
  \label{eq:4}
  \delta_Q L_0=\partial_\mu k^\mu_{Q} \iff Q^i\frac{\delta^L L_0}{\delta\phi^i}+\partial_\mu j^\mu_{Q}=0,\quad j^\mu_{Q}=\theta^\mu_Q-k^\mu_{Q}. 
\end{equation}
Here, $j^\mu_{Q}$ is the ``canonical'' representative for the Noether current. The
conservation law defines the Noether current up to the divergence of an
arbitrary antisymmetric tensor depending on the fields and their derivatives,
\begin{equation}
  \label{eq:Amb}
  j^\mu_{Q} \rightarrow  j^\mu_{Q} + \partial_\nu A^{\mu \nu} \, , \qquad A^{\mu \nu} = - A^{\nu \mu} \, .
\end{equation}
This is a consequence of the so-called ``algebraic Poincar\'e lemma'', which has
been derived and rederived independently by many authors (see \cite{Brandt:1989gy,Wald:1990mme,Dubois-Violette:1991dyw} for a sample
of different proofs from the physics literature; see the review \cite{Barnich:2000zw} for a longer
(but incomplete) list of references).

A useful property of global symmetries (in the sense of~\eqref{eq:4}) is that they are
automatically symmetries of the equations of motion (or in other words, the
equations of motion are covariant),
\begin{equation}
  \label{eq:33}
  \delta_Q \frac{\delta^L L_0}{\delta\phi^i}=-(-)^{iQ}(-\partial)_{(\lambda)}\Big[\frac{\partial Q^j}{\partial \phi_{(\lambda)}}
  \frac{\delta^Lf}{\delta\phi^j}\Big]\approx 0,
\end{equation}
where the sign factor involves a shorthand notation for the product of the
Grassmann parity of the field $\phi^i$ times that of $\delta_Q$. This follows
from formula~\eqref{eq:1} and the fact that Euler-Lagrange derivatives annihilate total
derivatives,
\begin{equation}
  \label{eq:34}
  \frac{\delta^L (\partial_\mu k^\mu )}{\delta\phi^i}=0. 
\end{equation}

{\bf Algebra}

Another useful property is that symmetries, which can be bosonic or fermionic,
form a graded Lie algebra. Indeed, using the fact that the graded commutator of
two variations is a variation,
\begin{equation}
  \label{eq:35}
  [\delta_{Q_1},\delta_{Q_2}]\equiv \delta_{Q_1}\delta_{Q_2}-(-)^{Q_1 Q_2}\delta_{Q_2}\delta_{Q_1}=\delta_{[Q_1,Q_2]},
\end{equation}
with
\begin{equation}
  \label{eq:36}
  [Q_1,Q_2]^i=\delta_{Q_1}Q^i_2-(-)^{Q_1 Q_2}\delta_{Q_2}Q^i_1, 
\end{equation}
one finds that the graded commutator of two symmetries leaves the Lagrangian
invariant up to a total derivative,
\begin{equation}
  \label{eq:37}
  \delta_{[Q_1,Q_2]}L_0=\partial_{\mu}(\delta_{Q_1}k^\mu_{Q_1}-(-)^{Q_1Q_2}
  \delta_{Q_2}k^\mu_{Q_1})\iff  [Q_1,Q_2]^i\frac{\delta^LL}{\delta \phi^i}+\partial_\mu j^\mu_{[Q_1,Q_2]}=0 \, ,
\end{equation}
i.e., is a symmetry. An explicit expression for the canonical representative of
the Noether current is given by
\begin{equation}
  \label{eq:38}
  j^\mu_{[Q_1,Q_2]}
=\theta^\mu_{[Q_1,Q_2]}-\delta_{Q_1}k^\mu_{Q_1}+(-)^{Q_1Q_2}\delta_{Q_2}k^\mu_{Q_1}.
\end{equation}

A well known fact of Hamiltonian dynamics is that the Poisson brackets of the
conserved charges associated with symmetries form a (projective) representation
of the algebra. Since the conserved charges are also the symmetry generators,
this Poisson bracket relation can be interpreted as expressing the variation of
one charge under the symmetry generated by the other in terms of the algebra.
The corresponding statement for the currents reads \cite{Dickey:1991}
\begin{equation}
  \label{eq:32}
  \delta_{Q_1} j^\mu_{Q_2}=j^\mu_{[Q_1,Q_2]}+\partial_{\nu}k^{[\mu\nu]}_{Q_1,Q_2}
  +S^\mu_{Q_1}(Q_2,\frac{\delta^LL_0}{\delta\phi^i})\approx j^\mu_{[Q_1,Q_2]}+\partial_{\nu}k^{[\mu\nu]}_{Q_1,Q_2}, 
\end{equation}
for some $k^{[\mu\nu]}_{Q_1,Q_2}$. For completeness, the proof of this important
relation, which plays a central role in the sequel, is given in Appendix {\bf
  \ref{sec:useful-formulas}}. Here, $S^\mu_{Q_1}(Q_2,\frac{\delta^LL_0}{\delta\phi^i})$ is an expression linear in
the Euler-Lagrange derivatives of $L_0$ and their spacetime derivatives, also
introduced in the appendix. The symbol $\approx$ means here ``equal when the equations
of motion derived from $S_0$ hold''. For simplicity, we have assumed here the
global validity of the algebraic Poincar\'e lemma so that no extension may
arise.

\subsection{Noether's second theorem}\label{sec:gauge-approach}

We now turn to Noether's second theorem, which deals with 
gauge transformations of the form
\begin{equation}
  \label{eq:18}
  \delta_{\epsilon} \phi^i=R^{i}_\alpha(\epsilon^\alpha)=R^{i(\mu)}_\alpha\partial_{(\mu)}\epsilon^\alpha=R^i_\alpha \epsilon^\alpha
  +R^{i\mu}_\alpha\partial_{\mu}\epsilon^\alpha+\dots,\forall \epsilon^\alpha(x), 
\end{equation}
where $\epsilon^\alpha$ are arbitrary infinitesimal spacetime functions. These
transformations are gauge symmetries if they leave the Lagrangian invariant up
to a total derivative, i.e.,
\begin{equation}
  \label{eq:20}
  \delta_{\epsilon} L_0=\partial_\mu k^\mu_{\epsilon} \iff
  R^i_\alpha(\epsilon^\alpha)\frac{\delta^L L_0}{\delta\phi^i}+\partial_\mu j^\mu_{\epsilon}=0,\quad j^\mu_{\epsilon}=\theta^\mu_{\epsilon}-k^\mu_{\epsilon},
\end{equation}
where $j^\mu_{\epsilon}$ is the canonical representative of the Noether current.

The main content of Noether's second theorem is that (i) the Euler-Lagrange
equations are not independent but obey ``Noether identities''; (ii) the Noether
current is trivial in the characteristic cohomology, i.e., equal on-shell to the
divergence of an antisymmetric tensor.

To derive these results, one performs integrations by parts on the
derivatives of the gauge parameter to show directly that
\begin{equation}
  \label{eq:5}
  R^i_\alpha(\epsilon^\alpha)\frac{\delta^L L_0}{\delta\phi^i}=(-)^{\alpha i}
   R^{+i}_\alpha(\frac{\delta^L L_0}{\delta\phi^i})\epsilon^\alpha+\partial_\mu S^\mu_{\epsilon},
\end{equation}
where
\begin{equation}
  \label{eq:6}
  \begin{split}
    & R^{+i}_\alpha(\frac{\delta^L L_0}{\delta\phi^i})=(-\partial)_{(\mu)}(R^{i(\mu)}_\alpha\frac{\delta^L L_0}{\delta\phi^i})
      =R^{i}_\alpha\frac{\delta^L L_0}{\delta\phi^i}
    -\partial_\mu(R^{i\mu}_\alpha\frac{\delta^L L_0}{\delta\phi^i})+\dots,\\
    & S^\mu_{\epsilon}=R^{i\mu}_\alpha\epsilon^\alpha\frac{\delta^L L_0}{\delta\phi^i}+\dotsc \approx 0.
  \end{split}
\end{equation}
Subtracting~\eqref{eq:5} from~\eqref{eq:20} yields 
\begin{equation}
  \label{eq:7}
  \partial_\mu j^\mu_\epsilon =-(-)^{\alpha i}
   R^{+i}_\alpha(\frac{\delta^L L_0}{\delta\phi^i})\epsilon^\alpha-\partial_\mu S^\mu_{\epsilon}.
\end{equation}
Replacing $\epsilon^\alpha$ by new independent (fermionic) fields $C^\alpha$ and taking
Euler-Lagrange derivatives with respect to $C^\alpha$, that kill total derivatives,
one gets the Noether identities
\begin{equation}
  \label{eq:2}
   (-)^{i(\alpha+1)}R^{+i}_\alpha(\frac{\delta^L L_0}{\delta\phi^i})=0,
 \end{equation}
 which is the first part of the theorem. [Since the equations (\ref{eq:7}) are linear
 in the $\epsilon^\alpha$'s, their Grassmann parity does not matter and their replacement by
 the ghosts is thus permissible. It is at this stage just a matter of
 convenience, performed in anticipation of the subsequent BRST analysis.]

 Using these identities in equation~\eqref{eq:5} then shows that $-S^\mu_C$ (defined by \eqref{eq:6}
 with the parameters $\epsilon^\alpha$ replaced by the ghosts) is a weakly vanishing
 representative for the Noether current associated to gauge symmetries,
 \begin{equation}
   \label{eq:29}
   R^i_\alpha(C^\alpha)\frac{\delta^L L_0}{\delta\phi^i}=\partial_\mu S^\mu_{C},
 \end{equation}
 while when used in~\eqref{eq:7} together with the algebraic Poincar\'e
 lemma,  one gets that the Noether current $j^\mu_C$ is trivial,
 \begin{equation}
   \label{eq:8}
   j^\mu_C=-S^\mu_C+\partial_\nu k^{[\mu\nu]}_C  \approx \partial_\nu k^{[\mu\nu]}_C, 
 \end{equation}
which is the second part of the theorem.

\section{Noether's 1.5 theorem}
\label{Sec:Noether15T}

\subsection{A quick review of the antified formalism}
The central object in the antifield (or BV-BRST) formalism \cite{Batalin:1981jr,Batalin:1983ggl,Batalin:1985qj} is the solution
of the master equation. We will only sketch here the ideas underlying the
general construction, referring to the original papers and to \cite{Henneaux:1989jq,Henneaux:1992ig} for more
information. We give more details for the irreducible case in Appendix {\bf
  \ref{App:BVIrr}}.

One extends the space of the original fields by introducing ghosts, ghosts for
ghosts if the symmetries are reducible, antighosts for gauge fixing (``non
minimal sector'') and the corresponding antifields. Let
$\varphi^A=(\phi^i,C^\alpha,\widebar C_\alpha,\dots)$ denote the fields, ghosts, antighosts, and
ghost for ghosts if any, and $\varphi_A^*$ denote the corresponding antifields. If we
denote $g_{(A)}$ the ghost number of the field $\vp^{A}$, the antifield $\varphi_A^*$
has ghost number $-g_{(A)}-1$. In particular, $(\phi^i,C^\alpha,\widebar C_\alpha)$ carry
ghost numbers $(0,1,-1)$.

Let $z \equiv (\varphi, \varphi^*)$.
For local functionals $F_{i}[z]=\int f_i\ d^n x$, the antibracket is defined by
\begin{equation}
  \label{eq:14}
  (F_1,F_2)=\int\, \big[\frac{\delta^R f_1}{\delta \varphi^A}\frac{\delta^L f_2}{\delta \varphi^*_A}
  - \frac{\delta^R f_1}{\delta \varphi^*_A}\frac{\delta^L f_2}{\delta \varphi^A}\big]. 
\end{equation}
Note that for any even functional $F$,
\begin{equation}
  \label{eq:23}
  \frac 12 (F,F)=\int \frac{\delta^R f}{\delta \varphi^A}\frac{\delta^L f}{\delta \varphi^*_A}\ d^n x
  =- \int \frac{\delta^R f}{\delta \varphi^*_A}\frac{\delta^L f}{\delta \varphi^A}\, d^n x. 
\end{equation}

The ghost-number-zero solution $S[\varphi, \varphi^*]$ of the master equation is obtained by
adding to the classical action $S_0[\phi]$ terms containing the ghosts and the
antifields, in such a way that $S$ fulfills the crucial master equation \be
\frac 12 (S,S) = 0 \, . \ee Besides $S_0$, which is requested to be equal to the
original gauge-invariant action, the first terms in the expansion of $S$ are
also given in terms of the gauge symmetries and the reducibility identities (if
any). Once the ``minimal solution'' is determined, one must include in addition
the ``non-minimal sector'' containing the antighosts, which is necessary when it
comes to fix the gauge.

The solution of the master solution $S[\varphi, \varphi^*]$ is obtained recursively starting
from $S_0$. One way to understand the construction is through ``Homological
Perturbation Theory'' (HPT), which provides also a geometric insight into the
classical BRST cohomology \cite{Fisch:1989rp}.

One crucial feature of $S[\varphi, \varphi^*]$ is that it is a local functional \cite{Henneaux:1990rx} (see
also \cite{Dubois-Violette:1991dyw}), \be S[\varphi, \varphi^*] = \int L \ d^n x \, , \ee where $L$ reduces to $L_0$ when
all the fields but the original ones are set to zero. By construction, this
action is invariant under the off-shell nilpotent BRST operator $s$ that acts on
fields and antifields as,
\begin{equation}
\label{def_s_BV}
s\varphi^A=(S,\varphi^A)=-\frac{\delta^R L}{\delta \varphi^*_A} \, ,\quad
  s\varphi^*_A=(S,\varphi^*_A)=\frac{\delta^R L}{\delta \varphi^A} \,  ,
\end{equation}
and extended to the derivatives of the fields in the standard way as before.
Furthermore, by introducing the operator $G$ whose eigenvalues are the ghost
numbers,
\begin{equation}
  \label{eq:56}
  G = g_{(A)} \partial_{(\mu)}\vp^A\frac{\partial^L}{\partial \vp^A_{(\mu)}}
  - (g_{(A)} + 1) \partial_{(\mu)}\vp^*_A\frac{\partial^L}{\partial \vp^*_{A,(\mu)}} \,    ,
\end{equation}
the fact that $S$ is of ghost number $0$ may be expressed in the form of a
symmetry as
\begin{equation}
  \label{eq:44}
  GS=0. 
\end{equation}

Gauge fixing is obtained by chosing a local ``gauge fixing fermion''
$\Psi[\varphi] = \psi d^nx$ of ghost number $-1$ and eliminating the antifields
through \be \varphi^*_A = \frac{\delta^L\psi}{\delta\varphi^A} \, . \ee The resulting gauge fixed action
is local
\begin{equation}
  \label{eq:55}
  S^g[\varphi]=S[\varphi,\frac{\delta^L\psi}{\delta\varphi}] = \int\, L^g\ d^nx,
\end{equation}
has ghost number zero and is invariant under the gauge fixed BRST operator $\gamma^g$,
\be
\gamma^g\varphi^A=( s \varphi^A)\Big\vert_{\varphi^*_A =  \tfrac{\delta^L\psi}{\delta\varphi^A}} \, .
\ee

The gauge fixed BRST operator $\gamma^g$ generically is only nilpotent on-shell for
the gauge fixed equations of motion,
\begin{equation}
  \label{eq:54}
  (\gamma^g)^2\approx 0 \, .
\end{equation}
One can define a gauge fixed BRST cohomology and show that it is equal to the
orifinal one \cite{Henneaux:1989jq} (see\cite{Barnich:1999cy,Barnich:2003tr} for further considerations on cohomological issues
related to locality).

\subsection{The $1.5$ theorem}

The gauge fixed BRST symmetry is a global symmetry of the local gauge fixed
action. We can apply to it Noether's first theorem. Let $j^{\mu}_{\gamma^g}$ be the
Noether current associated to BRST invariance of the gauge fixed action,
\begin{equation}
  \label{eq:58}
  \gamma^g\varphi^A\frac{\delta^L L^g}{\delta\varphi^A}+\partial_\mu j^{\mu}_{\gamma^g} =0.
\end{equation}

The gauge fixed action, which has ghost number zero, is also invariant under the
ghost number symmetry, $G L^g=0$. We denote by $j^{g\mu}_G$ a representative for
the Noether current of the ghost number symmetry,
\begin{equation}
  \label{eq:57}
  G L^g=0\iff g_{(A)} \vp^{A}\frac{\delta^L L^g}{\delta \vp^{A}} +\partial_\mu j^{g\mu}_G=0.  
\end{equation}

Now, the BRST transformation and the ghost number symmetry form the algebra \be
[G,\gamma^g]=\gamma^g \, , \ee which expresses the fact that $\gamma^g$ raises the ghost number
by $1$. Using~\eqref{eq:32} in the particular case where $\delta_{Q_2}$ is the ghost number
operator $G$ and $j^\mu_{Q_2}$ its Noether current $j^{g\mu}_G$, while $\delta_{Q_1}$ is
$\gamma^g$, one then gets
\begin{equation}
  \label{eq:59}
  \gamma^g j^{g\mu}_G\approx -j^\mu_{\gamma^g}+\partial_\nu k^{[\mu\nu]}_{\gamma^g,G} \, .
\end{equation}
This is precisely  Noether's 1.5 theorem
\begin{equation}
  \label{eq:59bis}
  \boxed{j^\mu_{\gamma^g}\approx -\gamma^g j^{g\mu}_G+\partial_\nu k^{[\mu\nu]}_{\gamma^g,G}}\, ,
\end{equation}
which expresses that modulo the (gauge fixed) equations of motion, the BRST
Noether current is BRST trivial up to the divergence of an antisymmetric tensor
\cite{Baulieu:2024oql,Baulieu:2024rfp}.

\section{BRST master current}
\label{Sec:MasterC}

Noether's 1.5 theorem establishes the triviality of the BRST current. This is
quite satisfactory since BRST invariance captures gauge invariance for which the
Noether current is trivial according to Noether's second theorem.

This suggests that another derivation more in line with Noether's second theorem
should exist. This different derivation would rely on the local gauge invariant
structure prior to gauge fixing. Such a derivation indeed exists and is the
object of the next two sections. The BRST master current is central in that
derivation.

\subsection{Definition and general properties  of BRST master current}\label{sec:noethers-1.5-theorem-1}

A local version of the master equation is
\begin{equation}
  \label{eq:24}
  \frac{\delta^R L}{\delta \varphi^A}\frac{\delta^L L}{\delta \varphi^*_A}=
  - \frac{\delta^R L}{\delta \varphi^*_A}\frac{\delta^L L}{\delta \varphi^A}=-\partial_\mu j^\mu_s, 
\end{equation}
where $j^\mu_s$ is by definition the BRST master current. It depends on the
fields, the antifields and their derivatives. Using \eqref{def_s_BV}, Eq.~\eqref{eq:24} may also by
written as
\begin{equation}
  \label{eq:25}
  (-)^As\varphi^As\varphi^*_A=s\varphi^A\frac{\delta^L L}{\delta\varphi^A}
  =\frac 12 sz^a\frac{\delta^L L}{\delta z^a}=-\partial_\mu j^\mu_s. 
\end{equation}
If we define $k^\mu_{s\varphi},\theta^\mu_{s\varphi}$ through
\begin{equation}
  \label{eq:41}
  \partial_{(\mu)}s\varphi^A\frac{\partial^L L}{\partial\varphi^A_{(\mu)}}=\partial_\mu k^\mu_{s\varphi},\quad
  \partial_{(\mu)}s\varphi^A\frac{\partial^L L}{\partial \varphi^A_{(\mu)}}
  =s\varphi^A\frac{\delta^L L}{\delta \varphi^A}+
  \partial_\mu \theta^\mu_{s\varphi},
\end{equation}
a canonical representative for the
master current is
\begin{equation}
  \label{eq:42}
  j^\mu_s=\theta^{\mu}_{s\varphi}-k^\mu_{s\varphi}. 
\end{equation}

Note that the master current receives no contribution from the non-minimal
sector associated with gauge fixing, because the variables of the non minimal
sector appear purely algebraically (with no derivatives) in the master equation.
We can thus restrict the sum in (\ref{eq:24}) or (\ref{eq:25}) to the minimal sector, which we
shall do until we get back to the problem of gauge fixation.

Let us denote by $j^{1,n-1}_s=j^\mu_s d^{n-1} x_\mu$ the ghost-number-one
$(n{-}1)$-form associated to the master current.

{\bf Claims:} (i) Up to the divergence of a superpotential, the master current
is an extension by terms of antifield number at least $1$ of the Noether current
$-S^\mu_C$ that was weakly vanishing on the original gauge invariant equations of
motion,
\begin{equation}
  \label{eq:26}
  j^\mu_s=-S^\mu_C+\partial_{\nu} k^{[\mu\nu]}_C+\dots.
\end{equation}

(ii) The master $(n{-}1)$-form is a solution to the consistency condition,
\begin{equation}
  \label{eq:27}
  s j^{1,n-1}_s+d j^{2,n-2}_s=0 \, ,
\end{equation}
for some $(n{-}2)$-form of ghost number $2$.

(iii) The master $(n{-}1)$-form is a trivial solution to the consistency condition,
\begin{equation}
  \label{eq:28}
  j^{1,n-1}_s=s \eta^{0,n-1}+d\eta^{1,n-2},
\end{equation}
for some $\eta^{0,n-1}$, $\eta^{1,n-2}$.

{\bf Proofs:}

(i) Equation~\eqref{eq:25} reads explicitly,
\begin{equation}
  \label{eq:30}
   (-)^A s\varphi^As\varphi^*_{A} =-\partial_\mu j^\mu_s.
\end{equation}
Projecting this equation on antifield number zero yields 
\begin{equation}
  \label{eq:30Bis}
  (-)^i \left.s\phi^i\right\vert_{0} \left.s\phi^*_{i}\right\vert_{0}
  =-\partial_\mu ( \left.j^\mu_s\right\vert_{0}),
\ee
 i.e., inserting the explicit expressions,
\begin{equation}
  \label{eq:31}
  R^i_\alpha(C^\alpha)\frac{\delta^L L_0}{\delta \phi^i}=-\partial_\mu ( \left.j^\mu_s\right\vert_{0}).
\end{equation}
Using~\eqref{eq:29} for the left hand side and the algebraic Poincar\'e lemma proves the claim. 

(ii) We may write~\eqref{eq:25} as
\begin{equation}
  \label{eq:10}
  (-)^A s\varphi^As\varphi^*_{A}\, d^n x - d j^{1,n-1}_s=0.
\end{equation}
Applying $s$ and using nilpotency together with the algebraic Poincar\'e lemma
yields~\eqref{eq:27}.

(iii) Since the weakly vanishing Noether current $S^{\mu}_C$ is $\delta$ exact,
\begin{equation}
  \label{eq:43}
  S^\mu_C=\delta[ - R^{i\mu}_\alpha C^\alpha \phi^*_i+\dots],
\end{equation}
where $\delta$ is the Koszul--Tate differential and where the dots now indicate
additional terms that come from multiple
integrations by parts. It follows that the term in antifield number $0$ of
$j^{1,n-1}_s$ is $\delta$ modulo $d$ exact. That this implies that $j^{1,n-1}_s$ is
$s$ modulo $d$ exact is a consequence of the results of \cite{Barnich:1994db}
(see theorem 7.1, equation 7.6 of~review \cite{Barnich:2000zw}).
\qed

\subsection{Explicit expression}\label{sec:noeth-1.5:-expl}

The proof of the previous claim uses again HPT with an expansion in antifields. One might thus wonder
whether it is possible to directly construct an expression for the $s$ exact
terms in~\eqref{eq:28} from the terms of the master action itself, which in turn have been
constructed by applying HPT using the same initial data. This is the object of
this subsection.

A local form of equation~\eqref{eq:44} is 
\begin{equation}
  \label{eq:49}
  G L=0\iff
  g_{(A)} \vp^{A}\frac{\delta^L L}{\delta \vp^{A}}  - (g_{(A)} + 1) \vp^*_A \frac{\delta^L L}{\delta \vp^*_A} 
  +\partial_\mu j^\mu_G=0 \,  ,
\end{equation}
where the canonical representative for the ghost number Noether current 
is $j^\mu_G=\theta^\mu_{Gz}$.

{\bf Claim:} (iv) Up to the divergence of a superpotential, the BRST master current is
the BRST variation of minus the Noether current $j^\mu_G$,
\begin{equation}
  \label{eq:47}
 \boxed{ j^\mu_s=-sj^\mu_G+\partial_\nu k^{[\mu\nu]}_s} \, .
\end{equation}

{\bf Proof:}

The idea is to apply a BRST variation to the equation (\ref{eq:49}).  A direct computation
using $s^2=0$ (which by \eqref{def_s_BV} implies that $ s\frac{\delta^L L}{\delta z^a}=0$) gives
\begin{equation}
  \label{eq:51}
   g_{(A)} s \vp^{A}\frac{\delta^L L}{\delta \vp^{A}}  - (g_{(A)} + 1) s \vp^*_A \frac{\delta^L L}{\delta \vp^*_A}   + \partial_\mu sj^\mu_G=0. 
\end{equation}
Using now that $s \vp^*_A \tfrac{\delta^L L}{\delta \vp^*_A} = s \vp^{A}\tfrac{\delta^L L}{\delta \vp^{A}}$, 
we get
\begin{equation}
  \label{eq:53}
  -s \vp^{A}\frac{\delta^L L}{\delta \vp^{A}}+\partial_\mu(s j^\mu_G)=0. 
\end{equation}
Subtracting this expression from the definition of the master current in~\eqref{eq:25} and using
the algebraic Poincar\'e lemma yields the result.

\section{Returning to Noether's 1.5 theorem}
\label{Sec:2nd15}

Equation (\ref{eq:47}) expressing the triviality of the BRST master current takes the same form as Eq.~\eqref{eq:59bis} expressing the triviality of the BRST Noether current.  However, it involves both the fields and the antifields and holds off-shell prior to any gauge fixing.  

One can easily connect the two by fixing the gauge in (\ref{eq:47}) .  To that end, as recalled in Appendix {\bf \ref{sec:gauge-fixat-thro}}, one first performs a canonical transformation (canonical in the antibracket, sometimes called anticanonical transformation).  This is easy to achieve because the master equation is invariant under canonical transformations. 

The canonical transformation relevant for gauge fixing is
\begin{equation}
  \label{eq:65bis}
  \varphi^A=\tilde\varphi^A,\quad \varphi^*_A=\tilde\varphi^*_A+\frac{\delta^L\psi}{\delta\varphi^A} \, ,
\end{equation}
where $\Psi[\varphi]= \int  \psi \, d^n x$ is the gauge fixing fermion.
Equation (\ref{eq:47}) then takes the same form in terms of the
new variables as it does in terms of the old ones.  So one has also
\begin{equation}
  \label{eq:70}
  \boxed{\tilde j^\mu_s=-\tilde s \tilde j^\mu_G+\partial_\nu \tilde k^{[\mu\nu]}_s} \, .
\end{equation}

We now fix the gauge by setting the shifted antifields $\tilde\varphi^*_A$ to zero. Doing this in~\eqref{eq:70} and using the formulas of Appendix {\bf \ref{sec:gauge-fixat-thro}} yields~\eqref{eq:59bis}.

This second derivation of Noether's $1.5$ theorem has the interest of connecting the BRST Noether currents appearing in the various gauge fixed versions of the theory to a single, gauge independent, master current.

\section{Conclusions}
\label{Sec:Conclusions}

In this paper, we have provided two different proofs of the Noether's $1.5$ theorem.  The first one is direct and consists in applying Noether's first theorem to the gauge fixed action, as well as in exploiting the consequences of the algebra of the Noether currents.  The second one uses gauge fixing only at the very last step and is closer to Noether's second theorem. It sheds furthermore interesting light on the structure of the BRST Noether current of the gauge fixed theory by relating it to the BRST master current involving both the fields and the antifields.  The ideas and methods of homological perturbation theory in the BV-BRST context \cite{Fisch:1989rp} (see also reviews \cite{Henneaux:1989jq,Henneaux:1992ig}) prove to be central.

The graded algebra of the (antifield dependent) BRST
symmetry $s$ and the ghost number symmetry plays a crucial role in the
constructive proof of the triviality of the master current. This reflects the
well-known fact that, on the level of the Hamiltonian BFV-BRST approach and in
operator quantization, the algebra of the BRST charge and the ghost number
generator is an important ingredient in the formalism.  Although a central term could in principle occur in the Poisson bracket algebra of the charges representing the BRST-ghost number algebra, none appears in our case since there is no central term of ghost number one.

Our considerations here are purely classical. Furthermore, boundary conditions have played no role so that the divergences of superpotentials that one may add in the context of Noether's first theorem were left undetermined. Non-trivial conserved $(n{-}2)$-forms may be related to ``global reducibility'' parameters, while asymptotic versions thereof arise when imposing boundary conditions and are directly related to asymptotic symmetries. In this context, suitably adapted path integral techniques and Ward identities become relevant in order to study infrared properties and their quantum corrections. Some of these aspects will be considered elsewhere.

\section*{Acknowledgments} 

The work of GB and MH is partially supported by FNRS-Belgium (convention IISN 4.4503.15).
The work of MH is additionally supported by research funds from the Solvay Family. 

\begin{appendix}

\section{Useful formulas}\label{sec:useful-formulas}

In this appendix, we prove formula (\ref{eq:32}), 
\begin{equation}
  \label{eq:App1}
  \delta_{Q_1} j^\mu_{Q_2}=j^\mu_{[Q_1,Q_2]}+\partial_{\nu}k^{[\mu\nu]}_{Q_1,Q_2}+S^\mu_{Q_1}(Q_2,\frac{\delta^LL_0}{\delta\phi^i})\approx j^\mu_{[Q_1,Q_2]}+\partial_{\nu}k^{[\mu\nu]}_{Q_1,Q_2} .
\end{equation}

Applying $\delta_{Q_1}$ to $Q^i_2\frac{\delta^LL_0}{\delta\phi^i}+\partial_\mu j^\mu_{Q_2}=0$ yields
\begin{equation}
  \label{eq:39}
  \delta_{Q_1}Q^i_2\frac{\delta^LL_0}{\delta\phi^i}+(-)^{Q_1(Q_2+i)} Q^i_2\delta_{Q_1}
  \frac{\delta^LL_0}{\delta\phi^i}+\partial_\mu \delta_{Q_1}j^\mu_{Q_2}=0 \, .
\end{equation}
Now the variations of the Euler-Lagrange derivatives can be expressed as
(equation~(5.86) of~\cite{Olver:1993} or equation~(6.43) of~\cite{Barnich:2000zw})
\begin{equation}
  \label{eq:1}
  \delta_{Q}\frac{\delta^L f}{\delta \phi^i}=(-)^{iQ}\frac{\delta^L }{\delta \phi^i}\Big[Q^j
  \frac{\delta^L f}{\delta \phi^j}\Big]-(-)^{iQ}(-\partial)_{(\lambda)}\Big[\frac{\partial Q^j}{\partial \phi^i_{(\lambda)}}\frac{\delta^Lf}{\delta\phi^j}\Big]. 
\end{equation}
If we apply this formula to the middle term of (\ref{eq:39}), by using that $Q_1$
defines a symmetry and that Euler-Lagrange derivatives annihilate total
derivatives, one finds after suitable integrations by parts
\begin{equation}
  \label{eq:40}
  \begin{split}
    (-)^{Q_1(Q_2+i)} Q^i_2\delta_{Q_1}\frac{\delta^LL_0}{\delta\phi^i} &
= -(-)^{Q_1(Q_2+i)}(-)^{iQ_1}Q_2^i(-\partial)_{(\lambda)}\Big[\frac{\partial Q^j_1}{\partial \phi^i_{(\lambda)}}\frac{\delta^LL_0}{\delta\phi^j}\Big]
 \\
 &=
  -(-)^{Q_1Q_2}\delta_{Q_2}Q_1^j\frac{\delta^LL_0}{\delta\phi^j}
  +\partial_\mu S^\mu_{Q_1}(Q_2,\frac{\delta^L L_0}{\delta\phi^i}),
  \end{split}
\end{equation}
where 
\begin{equation}
  S^\mu_{Q_1}(Q_2,\frac{\delta^L L_0}{\delta\phi})=(-)^{Q_1Q_2}Q^i_2\frac{\partial Q^j_1}{\partial \phi^i_\mu}\frac{\delta^LL_0}{\delta\phi^j}+\dots \, .
\end{equation}
Plugging this in \eqref{eq:39}, taking into account~\eqref{eq:37} and using again the
algebraic Poincar\'e lemma then leads to the desired result.

\section{Antifield formalism for irreducible gauge theories}\label{sec:bv-irreducible-gauge}

\label{App:BVIrr}

We give here explicit formulas for the BV-BRST construction \cite{Batalin:1981jr,Batalin:1983ggl,Batalin:1985qj} in the case of
irreducible theories.

\subsection{Master equation and solution}

We start with the gauge transformations~\eqref{eq:18}, which are assumed here to form an
irreducible, generating set of gauge transformations. In this case, one
considers the fields and ghosts $\varphi^A=(\phi^i,C^\alpha)$, together with their antifields
$z^a=(\varphi^A,\varphi^*_A)$. No ghost for ghost is necessary.

Let us denote by $ \delta_z = (\delta_\phi,\delta_{\phi^*},\delta_{C},\delta_{C^*})$ the operators whose
eigenvalues are the number of fields $\phi^i,C^\alpha,\phi^*_i,C^*_\alpha$ and their
derivatives,
\begin{equation}
  \label{eq:11}
  \delta_z=\partial_{(\mu)} z^a\frac{\partial^L}{\partial z^a_{(\mu)}} \, . 
\end{equation}
Three gradings are relevant here, the ghost number $g$, the pure ghost number
$pg$, and the antifield number $a$,  defined respectively 
as the eigenvalues of the operators
\begin{equation}
  \label{eq:12}
G=  \delta_C-\delta_{\phi^*}-2\delta_{C^*},\quad PG=\delta_C, \quad A=\delta_{\phi^*}+2\delta_{C^*}, 
\end{equation}
where
\begin{equation}
  \label{eq:63}
  G=PG-A,\quad PG(\varphi^*_A)=0,\quad A(\varphi^A)=0,\quad G(\varphi^*_A)=-G(\varphi^A)-1. 
\end{equation}

The minimal BV master action takes the form
\begin{equation}
  \label{eq:9}
  S[z] =\int L\ d^n x, \quad L=\sum_{k\geq 0} L_k=L_0-R^i_\alpha(C^\alpha)\phi^*_i+L_{k\geq 2}, 
\end{equation}
and can be constructed as the ghost number $0$ solution to the master equation
\begin{equation}
  \label{eq:13}
  \frac 12 (S,S)=0, 
\end{equation}
using Homological Perturbation Theory, the resolution degree being the antifield
number denoted by a subscript in (\ref{eq:9}) \cite{Fisch:1989rp}. For Yang--Mills theories and
gravity, the integrand is linear in the antifields (``rank $1$ theories'') and of
the form
\begin{equation}
  \label{eq:15}
  L=L_0-R^i_\alpha(C^\alpha)\phi^*_i+\frac 12 f^\alpha_{\beta\gamma}(C^\beta, C^\gamma) C^*_\alpha .
\end{equation}

The nilpotent (antifield-dependent) BRST transformations are explicitly given
by,
\begin{equation}
  \label{eq:22}
  \begin{split}
  & s\phi^i=R^i_\alpha(C^\alpha)+\dots,\quad s C^\alpha=-  \frac 12 f^\alpha_{\beta\gamma}(C^\beta,C^\gamma)+\dots,\\
    & s\phi^*_i=\frac{\delta^R L_0}{\delta \phi^i}+\dots,\quad sC^*_\alpha=-R^{+i}_\alpha(\phi^*_i)(-)^{\alpha(i+1)}+\dots, 
  \end{split}
\end{equation}
where the dots denote terms of higher antifield number.

We note that as a direct consequence of~\eqref{def_s_BV} and of the nilpotentcy of
$s$, the BV variational derivatives are invariant and not merely covariant,
\begin{equation}
  \label{eq:46}
  s\frac{\delta^L L}{\delta z^a}=0=s\frac{\delta^R L}{\delta z^a}. 
\end{equation}

The BRST master current directly follows from these formulas. For theories of
rank $1$ of the form~\eqref{eq:15} with at most first order derivatives on the ghosts
$C^\alpha$, one has
\begin{equation}
  \label{eq:48}
  \begin{split}
    & R^i_\alpha(C^\alpha)=R^i_\alpha C^\alpha+R^{i\mu}_\alpha \partial_\mu C^\alpha,\\
    & \frac 12 f^\alpha_{\beta\gamma}(C^\beta,C^\gamma)=\frac 12 f^\alpha_{\beta\gamma}C^\beta C^\gamma+f^{\alpha\mu}_{\beta\gamma}\partial_\mu C^\beta C^\gamma,
  \end{split}
\end{equation}
and therefore
\begin{equation}
  \label{eq:45}
  j^\mu_G=C^\alpha\frac{\partial^L L}{\partial \pa_\mu C^\alpha}= -R^{i\mu}_\alpha C^\alpha \phi^*_i
  +f^{\alpha\mu}_{\beta\gamma} C^\beta C^\gamma C^*_\alpha. 
\end{equation}
In Yang--Mills, and Chern--Simons theory, only the first term is
present since the gauge algebra does not involve derivatives of the gauge
parameters, while in general relativity in metric formulation, there is also the
second term since the Lie bracket of vector fields involves derivatives of
vector fields.

\section{Gauge fixation}
\label{sec:gauge-fixat-thro}

To fix the gauge, one introduces the non-minimal sector. It is here that the
antighosts (and antighosts for antighosts in the reducible case) appear.

In the irreducible case, the standard non-minimal sector contains the fields
$\widebar C^\alpha, B^\alpha$ (antighosts and $B$-field) and their antifields
$\widebar C^*_\alpha,B^*_\alpha$. The antibracket is extended to include those fields and
antifields while the minimal BV master action is extended as
\begin{equation}
  \label{eq:60}
 S^{\mathrm{nm}}[z]=\int \,  L^{\mathrm{nm}} \ d^n x   ,\quad L^{\mathrm{nm}}=L- B^\alpha \widebar C^*_\alpha ,
\end{equation}
so that the BRST transformation in the non minimal sector takes the form of
trivial pairs that do not contribute to (local, antifield dependent) BRST
cohomology,
\begin{equation}
  \label{eq:61}
  s \widebar C^\alpha= B^\alpha,\quad s B^\alpha=0,\quad s B^*_\alpha=-\widebar C^*_\alpha,\quad s \widebar C^*_\alpha=0.  
\end{equation}
In the reducible case, antighosts for antighosts and additional auxiliary fields
must be introduced \cite{Batalin:1983ggl} (see also monograph \cite{Henneaux:1992ig}).

The associated extensions of the various ghost, pure ghost and antifield numbers
are
\begin{equation}
  \label{eq:62}
  G^{\mathrm{nm}}=G-\delta_{\bar C}-\delta_{B^*},\quad PG^{\mathrm{nm}}=PG-\delta_{\bar C},\quad A^{\mathrm{nm}}=A+\delta_{B^*}.
\end{equation}
Note that both the BRST and the ghost number currents are identical to those of
the minimal master action
\begin{equation}
  \label{eq:69}
  j^{\mathrm{nm}\mu}_s= j^\mu_s,\quad j^{\mathrm{nm}\mu}_G=j^\mu_G \, ,
\end{equation}
because the non-minimal sector is algebraic.

Gauge fixation is done through a canonical transformation
$(\varphi^A,\varphi^*_A)\to (\tilde\varphi^A,\tilde\varphi^*_A)$ modifying only the antifields, with a
generating functional that involves a gauge fixing fermion of ghost number $-1$
of the form
\begin{equation}
  \label{eq:66}
  \Psi[\varphi]=\int \,  \psi  \ d^n x \,  ,
\end{equation}
so that
\begin{equation}
  \label{eq:65}
  \varphi^A=\tilde\varphi^A,\quad \varphi^*_A=\tilde\varphi^*_A+\frac{\delta^L\psi}{\delta\varphi^A}. 
\end{equation}
Since the canonical transformation does not modify the fields, one might
drop all the tildes on the fields, but not on the antifields.

Because the transformation is canonical in the antibracket, if
$\tilde F(\tilde z)=F(z(\tilde z))$ and $(\cdot,\cdot\tilde)$ denotes the antibracket in
terms of $\tilde z$, then
\begin{equation}
  \label{eq:64}
  (\tilde F_1,\tilde F_2\tilde)=\widetilde{(F_1,F_2)}.
\end{equation}
In particular, if
\begin{equation}
  \label{eq:67}
  S^{\mathrm{gf}}[\tilde z]=\tilde S^{\mathrm{nm}},
\end{equation}
it follows that
\begin{equation}
  \label{eq:68}
  \frac{1}{2}(S^{\mathrm{gf}},S^{\mathrm{gf}}\tilde)=0,
  \quad \tilde s=(S^{\mathrm{gf}},\cdot\tilde),\quad \tilde s^2=0,
\end{equation}
so that the BRST differential $\tilde s$ that acts on, and involves, both
$\tilde\varphi^A,\tilde\varphi^*_A$ is nilpotent off-shell.

Let
\begin{equation}
  \label{eq:71} S^g[\varphi]=S^{\mathrm{gf}}[\varphi,\tilde\varphi^*=0]
  =S^{\mathrm{nm}}[\varphi,\frac{\delta^L\psi}{\delta\varphi}],
\end{equation}
be the gauge fixed BRST action considered in \eqref{eq:55} with the shifted antifields
$\tilde\varphi^*_A$ set to zero and consider the gauge fixed antifield number
\begin{equation}
  \label{eq:73}
  A^g=\delta_{\tilde\varphi^*} ,
\end{equation}
whose eigenvalues are one for each for of the antifields $\tilde\varphi^*_A$.
Decomposing $\tilde s$ according to this degree, one has
\begin{equation}
  \label{eq:74}
  \tilde s \varphi^A=\gamma^g\varphi^A+\lambda^g\varphi^A+\dots,\quad
  \tilde s\tilde \varphi^*_A=\delta^g\tilde \varphi^*_A+\gamma^g\tilde \varphi^*_A+
  \dots,
\end{equation}
where 
\begin{equation}
  \label{eq:75}
  \gamma^g\varphi^A=(\tilde s \varphi^A)|_{\tilde\varphi^*_A=0},\quad \delta^g\tilde\varphi^*_A
  =(\tilde s \tilde \varphi^*_A)|_{\tilde\varphi^*_A=0}=\frac{\delta^R L^g}{\delta\varphi^A},
\end{equation}
with $\gamma^g$ the gauge fixed BRST differential without antifields, $\delta^g$ the Koszul--Tate
differential for the gauge fixed stationary surface, while
$\lambda^g\varphi^A,\gamma^g\tilde \varphi^*_A$ denotes the terms of gauge fixed
antifield number $1$ and the dots denote terms of gauge fixed antifield
number strictly higher than $1$.

One has
\be
\gamma^g S^g[\varphi] = 0 \, , \qquad (\gamma^g)^2 \approx 0 \, ,
\ee
where $\approx$ means here ``equal modulo the gauge fixed equations of motion'' \cite{Henneaux:1989jq,Henneaux:1992ig}.
The operator $\gamma^g$ is the gauge fixed BRST operator of
Section~\ref{sec:noeth-first-theor}.

\end{appendix}

\end{document}